\documentclass[prx,twocolumn,showpacs,superscriptaddress,nofootinbib]{revtex4-1}
\usepackage{amsmath,amssymb}
\usepackage{bm}
\usepackage{graphicx}
\usepackage[colorlinks,citecolor=blue,linkcolor=red]{hyperref}
\usepackage{color}

\begin{document}
\preprint{LA-UR-12-24050}

\title{LDA+DMFT Approach to Magnetocrystalline Anisotropy of Strong Magnets}

\author{Jian-Xin Zhu} 
\email{jxzhu@lanl.gov}
\affiliation{Los Alamos National Laboratory, Los Alamos, New Mexico 87545, USA}

\author{Marc Janoschek}
\affiliation{Los Alamos National Laboratory, Los Alamos, New Mexico 87545, USA}

\author{Richard Rosenberg}
\affiliation{Advanced Photon Source, Argonne National Laboratory, Argonne, Illinois 60439, USA}

\author{Filip  Ronning}
\affiliation{Los Alamos National Laboratory, Los Alamos, New Mexico 87545, USA}

\author{J.~D.~Thompson}
\affiliation{Los Alamos National Laboratory, Los Alamos, New Mexico 87545, USA}

\author{Michael A. Torrez}
\affiliation{Los Alamos National Laboratory, Los Alamos, New Mexico 87545, USA}

\author{Eric D. Bauer}
\affiliation{Los Alamos National Laboratory, Los Alamos, New Mexico 87545, USA}

\author{Cristian~D.~Batista}
\email{cdb@lanl.gov}
\affiliation{Los Alamos National Laboratory, Los Alamos, New Mexico 87545, USA}

\begin{abstract}
The new challenges posed by the need of finding strong rare-earth-free magnets demand methods that can  predict magnetization and magnetocrystalline anisotropy energy (MAE). We argue that correlated electron effects, which are normally underestimated in band structure calculations, play a crucial role in the development of the orbital component of the magnetic moments.  
Because magnetic anisotropy arises from this orbital component, the ability to include correlation effects  has profound consequences on our predictive power of the MAE of strong magnets. Here we show that incorporating the local effects of electronic correlations with dynamical mean-field theory  provides reliable estimates of the orbital moment, the mass enhancement and the MAE of YCo$_5$.
\end{abstract}

\maketitle

\section{Introduction}
Magnets play a central role in different types of devices and motors, which are  at the heart of modern technology. 
There is an increasing need of permanent magnetic materials for energy conversion and power generation~\cite{Lewis12}.  
Magnetocrystalline anisotropy (MA) is one of the most important properties of permanent magnets~\cite{Kirchmayr79}.
Large MA  is  achieved in existing strong magnets by using rare-earth transition-metal intermetallic compounds, such as   SmCo$_5$ and Nd$_2$Fe$_{14}$B, which are of direct technological use.
However, the shortage of rare-earth elements  has triggered the search for rare-earth-free magnetic materials harnessing sources of magnetic anisotropy other than that provided by the rare-earth components~\cite{Lewis12}.
In order to guide this search, it is necessary to develop theoretical methods that can estimate the 
magnetocrystalline anisotropy energy (MAE) of $3d$, $4d$ and $5d$ transition metals, which are the  natural candidates for replacing 
rare-earth elements.

The contribution  of itinerant ferromagnetic electrons to MA arises from the spin-orbit (SO) interaction that couples the spin
and orbital components of the magnetic moments~\cite{JHvanVleck:1937}. MA results from the orbital component of the moment, which is sensitive 
to the lattice anisotropy.  The very first electronic structure analysis of MAE for Ni was conducted by Kondorskii and Straub~\cite{EIKondorskii:1972}.
While a band picture may provide a MAE of the right order of magnitude for certain transition metal ferromagnets~\cite{Bloch31,Brooks40},
accurate electronic structure calculations  of the MAE of 3$d$ metals,  such as Fe, Co and Ni, give numbers that are in disagreement with experiment~\cite{Daalderop90,JTrygg:1995}. Moreover, 
the wrong easy-axis is obtained for Ni. This failure has been attributed to either the omission of the orbital correlation induced by the intra-atomic Coulomb interaction between electrons~\cite{Jansen90} or the  limitation of band structure calculations for calculating energy differences of the order of 0.1~meV~\cite{Daalderop90}. 


YCo$_5$ has one of the largest MAEs among ferromagnets that do not include $f$-electron (actinide or lanthanide) ions. 
The MAE is more than 50 times larger than in the pure cobalt metal. Like SmCo$_5$,  it has an easy-axis parallel to the $c$-axis of its hexagonal lattice structure. The primitive unit cell contains six atoms with two different
cobalt sites, Co$_{\rm I}$ (2c) and Co$_{\rm II}$ (3g)~\cite{Nordstrom92}. Neutron scattering experiments 
by Schweizer {\it et al.} have reported  unusually large orbital moments on these Co sites~\cite{Schweizer80}: m$_{\rm orb}$(Co(2c))=0.46$\mu_B$ and 
m$_{\rm orb}$(Co(3g))=0.28$\mu_B$.  However, our x-ray magnetic circular dichroism (XMCD) measurements indicate that the average orbital moment of Co is $0.2$~$\mu_B$, in better agreement with the value of $0.25$~$\mu_B$ reported by Heidemann {\it et al.}~\cite{Heidemann75}. 
These measurements  suggest that the rather large  orbital magnetic moments of the Co atoms  are partially responsible for the strong MAE of YCo$_5$. Consequently, reliable estimates of the MAE require an accurate calculation of these orbital moments. This is not only true for YCo$_5$, but also  for any other strong magnet based on transition metals. We will then  use YCo$_5$ as a prototype compound  for developing and testing methods for calculating the orbital moments and the MAE of strong magnets.

Nordstr{\"o}m {\it et al.}~\cite{Nordstrom92}  have applied the force theorem to compute the MAE of YCo$_5$ from first-principles calculations.  It was found that,  in the absence of atomic orbital correlation, the MAE is too small and it even has the incorrect sign, in agreement with Ref.~\cite{Daalderop90}. After including the orbital polarization (OP) scheme suggested by Brooks~\cite{Brooks85,Eriksson90},  they were able to obtain a MAE that has the correct sign. However, the MAE value of about 50$\mu$Ryd, when extrapolated to the infinitesimal grid in the momentum space  although,  is still too small in comparison with the experimental value  of 292 $\mu$Ryd~\cite{Alameda80}.
A similar improvement is obtained for  estimations of the orbital magnetic moments of both Co sites. In absence of orbital correlation, the result is m$_{\rm orb}$(Co(2c))=0.1$\mu_B$ and 
m$_{\rm orb}$(Co(3g))=0.13$\mu_B$~\cite{Takahashi87}, while the inclusion of OP leads to m$_{\rm orb}$(Co(2c))=0.27$\mu_B$ and 
m$_{\rm orb}$(Co(3g))=0.20$\mu_B$~\cite{Nordstrom92}. 

The OP scheme is taken from the theory of open shell atoms within the Russel-Saunders 
coupling. The ground state energy gain, that is obtained by maximizing the orbital angular momentum $L$, is approximated by $E_{\text{OP}}=-B L^2/2$, where $B$ is the Racah parameter for $d$ sates.  This effect is just a consequence of the Coulomb interaction between $d$-electrons that occupy the same ion and it  must influence the final value of the orbital magnetic moment and the MAE.   However, it is well known that the on-site electron-electron Coulomb interaction also renormalize the band states (electrons tend to avoid each other), for which heavy fermion behavior in $f$-electron systems is a prototypical example~\cite{ACHewson:1993}.
Therefore, it is reasonable to expect that this second consequence of the Coulomb interaction will also affect the magnitude of the orbital magnetic moment and the MAE. Here we propose a method for including these additional correlations.

\section{Role of Coulomb Interaction on Local Moment Formation}
The effect of electron-electron interaction is to reduce the bandwidth of the quasi-particles and produce an incoherent component in their spectral weight. The most dramatic effect of this Coulomb repulsion is the emergence of Mott insulators in half-filled bands via localization of individual  electrons in their atomic orbitals. The electronic localization is accompanied by the formation of a local magnetic moment, whose spin and orbital components can be of the order of a Bohr magneton ($\mu_B$)~\cite{Goodenough63}. It is clear that Coulomb interaction cannot localize the electronic charge  away from half-filling. However, the band narrowing effect can be interpreted as a tendency towards localization that favors local moment formation. 
This simple reasoning suggest that the inclusion of electronic correlations should lead to more realistic values of the effective mass of the quasiparticles, orbital magnetic moments and MAE.  

Standard LDA calculations lead to orbital magnetic moments of order 0.1$\mu_B$. This result can be understood in the following way.  The typical bandwidths, $W$, of $3d$ metals like Fe or Co are of the order  of a few electron volts. The SO interaction is about $\lambda \simeq 0.05-0.07$ eV. In the absence of SO coupling, the ground state has zero orbital angular momentum, even if it has a net spin magnetization, because single-particle states with opposite values of the orbital magnetic moment are degenerate and therefore equally occupied.  A finite SO coupling term splits states with opposite values of orbital moment by an amount that is of order  $\lambda$. This observation implies that only the electronic states that are  within a distance $\lambda$ from the Fermi level contribute to orbital polarization. The fraction of electrons occupying these states is of order 
 $\lambda / W \simeq 0.02$. Because the maximum possible value of the orbital moment per atom is of order $1\mu_B$, this rough estimate indicates that  $m_{\rm orb} \lesssim 0.1~\mu_B$  in agreement with previous results from standard  band structure calculations \cite{Nordstrom92}. However, as  pointed out in the introduction, the orbital magnetic  moment of strong magnets, such as YCo$_5$, can be  higher than this rough estimate. 

It is natural to assume that the  discrepancy arises from the effects of rather strong intra and inter-atomic electronic correlations induced by the Coulomb interaction. The  improvement that is obtained after including the intra-atomic OP effect provides empirical support for this assumption. However, the most basic and general argument in favor of this assumption is 
that  intra-atomic Coulomb repulsion favors  local moment formation by suppressing double occupancy of single atomic orbitals.  The importance of correlation effects on the magnetic anisotropy of  Fe and Ni was already recognized more than ten years  ago by Yang {\it et al.}~\cite{Yang01}.
This problem is now timely because of the increasing need of finding strong magnets that are free of rare earth  elements.
Therefore, it is crucial to propose new methods that can incorporate the subtle effects of correlations in solids (the OP effect that we discussed above is already captured at the level of single-atom physics).  For this purpose we propose a method based on 
the combination of the dynamical mean-field theory (DMFT)  and the LDA~\cite{Kotliar06}. A similar approach has been successfully applied to the calculation of neutron magnetic form factors of actinides by applying an external magnetic field~\cite{Pezzoli11}, as well as the bulk and surface quasiparticle spectra~\cite{AGrechnev:2007} and the orbital magnetism~\cite{SChadov:2008} in Fe, Co, and Ni metals.
The basic idea is to treat each Co ion as an effective impurity that is embedded  into the bath generated by the rest of the ions. The single-ion interactions (including the OP) are captured by the single-impurity Hamiltonian. The correlations developed via the interplay between the single-ion terms  and the interaction with the bath (solid) are captured by a self-consistent treatment of the full Hamiltonian that we describe in the next section.

\section{Local Density Approximation Plus Dynamical Mean-Field Theory}
To study the role of electronic correlations on the orbital moment of the magnetic $3d$ ions by combining the LDA with dynamical mean-field theory (LDA+DMFT)~\cite{Kotliar06} we start with a generalized many-body Hamiltonian: 
\begin{eqnarray}
\hat{H} &=& \sum_{\mathbf{k},lm_l\sigma,l^{\prime}m_l^{\prime}\sigma^{\prime}} [H^{(0)}_{\mathbf{k}}]_{l m_l \sigma,l^{\prime} m_l^{\prime}\sigma^{\prime}} c_{\mathbf{k}lm_l\sigma}^{\dagger} c_{\mathbf{k}l^{\prime} m_l^{\prime} \sigma^{\prime}  }
\nonumber \\
&+& \frac{1}{2}  \!\!\!\! \sum_{\begin{subarray} i,l=2(3),m_l m^{\prime}_l, \\ m^{\prime\prime}_l m^{\prime\prime\prime}_l \sigma \sigma^{\prime} \end{subarray}}
\!\!\!\!\!\!\!\!
V_{m_l m^{\prime}_l m^{\prime\prime}_l m^{\prime\prime\prime}_{l} }
c_{ilm_l \sigma}^{\dagger} c_{ilm^{\prime}_{l} \sigma^{\prime}}^{\dagger} c_{ilm^{\prime\prime\prime}_{l} \sigma^{\prime}} c_{ilm^{\prime\prime}_{l}\sigma}.
\label{EQ:Hamil}
\end{eqnarray}
Here $\mathbf{k}$ is a wave vector of the Brillouin-zone, $i$ is a lattice site index for atoms with correlated orbitals, $l$ is the orbital angular momentum, $m_l=-l,-l+1,\dots,l-1,l$, and $\sigma$ is the spin projection quantum number. The field operator $c_{ilm_l\sigma}^{\dagger}$ ($c_{ilm_l\sigma}$) creates (annihilates) an electron with spin $\sigma$ and orbital indices $(l m_l)$ at site $i$, while  $c_{\mathbf{k}lm_l\sigma}^{\dagger}$ ($c_{\mathbf{k}lm_l\sigma}$) is the corresponding operator in  momentum space. The first term of $\hat{H} $ contains  the single-particle contribution, which is determined by solving the Kohn-Sham quasi-particle equations~\cite{RMMartin2004} within LDA. We note that the SO coupling can be included in a second variational way in the LDA Hamiltonian.  In the second term of $\hat{H} $ we restrict the Coulomb repulsion to the correlated orbitals (e.g., open shell Co 3$d$ orbitals $(l=2)$ or Ce 4$f$ orbitals $(l=3)$)  to reduce the complexity of the problem~\cite{Kotliar06}.  
The Coulomb matrix elements are obtained from atomic physics:
\begin{equation}
V_{m_l m^{\prime}_l m^{\prime\prime}_l m^{\prime\prime\prime}_l} = \sum_{k(even)=0}^{2l} a_{k}(m_l,m_l^{\prime},m_{l}^{\prime\prime},m_{l}^{\prime\prime\prime})F^{k} \;,
\end{equation}
where $F^{k}$ and $a_{k}$ are the Slater integrals and the corresponding expansion coefficients~\cite{Cowan1981}.  For solids, we identify the atomic Slater integral $F^{0}$ with the screened effective Coulomb interaction parameter $U$ of the correlated orbitals. As a common practice, higher order Slater integrals are reduced by 20\% from the atomic Hartree-Fock calculations due to screening effects~\cite{Czyzyk1994}. 

\begin{figure}
\includegraphics[width=1\linewidth]{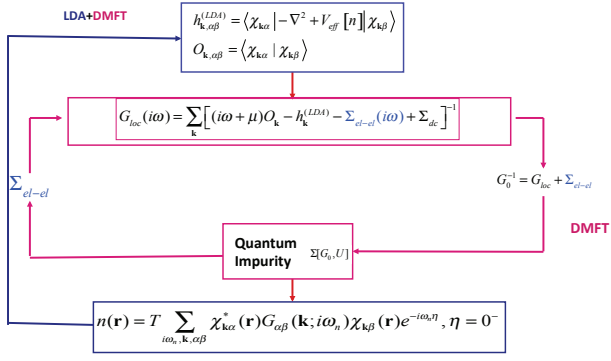}
\caption{\label{Fig1} Schematic description of the DMFT+LDA approach. }
\end{figure}

Within DMFT, the lattice problem of Eq.~(\ref{EQ:Hamil}) is mapped onto a 
multi-orbital quantum single impurity problem subject to the self-consistency condition (see Fig.~\ref{Fig1}):
\begin{equation} 
{\hat {\mathcal G}}^{-1}(i\omega_{n}) = {\hat G}_{\rm loc}^{-1} (i \omega_n) + {\hat \Sigma} (i\omega_{n})\;.
\end{equation}
Here $\hat{\mathcal{G}}(i\omega_{n})$ is the
Weiss function, $\hat{\Sigma}(i\omega_{n})$ is a
$\mathbf{k}$-independent self-energy, and the local Green's
function is defined as ${\hat G}_{\rm loc}(i\omega_{n})
=\sum_{\mathbf{k}}\hat{G}_{\mathbf{k}}(i\omega_{n})/N$, where
the lattice Green's function reads
\begin{equation}
\hat{G}_{\mathbf{k}}(i\omega_{n}) = [(i\omega_{n} + \mu)\hat{I} - \hat{H}^{0}(\mathbf{k}) -
\hat{\Sigma}(i\omega_{n})]^{-1}\;.
\end{equation}
$\hat{I}$ is the identity matrix in the complete tight-binding basis and $\mu$ is the chemical
potential. Because  we have  added the on-site Coulomb terms  to the  correlated valence orbitals only, it is evident that the self-energy $\hat{\Sigma}$ matrix has nonzero elements only within the $10\times 10$ $d$-$d$ block for the case of  valence $d$-orbitals, or the  $14\times 14$ $f$-$f$ block for the case of valence $f$-orbitals.  This self-energy matrix is a function of  the Matsubara frequency:  
$\Sigma^{dd(ff)}_{m_l \sigma, m_{l}^{\prime} \sigma^{\prime}} (i\omega_{n})$. 
Correspondingly, the local Green's function for the correlated orbitals has the same structure
$G^{dd(ff)}_{{\rm loc},m_{l} \sigma,m_l^{\prime}\sigma^{\prime}}(i\omega_{n})$.
We assume that the dominant contributions to the spin and orbital components of the magnetic moments come from the correlated orbitals (Co 3$d$-orbitals for the case of YCo$_5$). After obtaining the local Green's function for the correlated orbitals through the full self-consistency, we can evaluate the spin and orbital moments by 
computing
$
M_{s}=\sum_{m_l\sigma}  \sigma \rho_{m_{l}\sigma,m_{l}\sigma}$
{\rm and}
$
M_{orb}=\sum_{m_l\sigma}   m_l \rho_{m_{l}\sigma,m_{l}\sigma},
$
respectively, in the spherical harmonics basis. Here the density matrix is related to the local Green's function as 
\begin{equation}
\hat{\rho} = \hat{G}^{dd(ff)}_{loc}(\tau\rightarrow 0^{-}) = \frac{1}{\beta} \sum_{i\omega_n} \hat{G}^{dd(ff)}_{loc}(i\omega_n)e^{-i\omega_n 0^{-}}\;, 
\end{equation}
where $\beta=1/k_{B}T$, with $k_{B}$ and $T$ the Boltzmann constant and temperature, respectively. 

In earlier applications of the LDA+DMFT method, it is common use to  rotate the local Green's function and the corresponding self-energy into a basis in which the diagonal matrix elements are dominant in order to neglect the off-diagonal elements. For example,  for actinide based materials, the correlated 5$f$ orbitals are rotated into the $J$-$J$ basis because of the dominant  SO coupling~\cite{JHShim:2007,JXZhu:2013}. In contrast, SO coupling is subdominant for $d$-electron materials, like transition metal oxides, and the self-energy and local Green's function matrices are  diagonal  in the 
crystal field basis when the SO coupling is 
neglected~\cite{VIAnisimov:1997,LCraco:2004,KHeld:2001,MSLaad:2003}. 
However,  the off-diagonal matrix elements cannot be neglected  if our goal is to  compute the MAE (the SO coupling must be included to obtain a finite MAE and the orbital magnetic moment has only off-diagonal contributions in this basis).  
This situation requires a further development of quantum impurity solvers to meet this challenge and similar challenges posed by other correlated electron materials, such as the inclusion of crystal field terms in 4$f$ and 5$f$ compounds.

\section{Computational Results}
Here we use the spin-polarized $T$-matrix fluctuation-exchange approximation technique (SPTF)~\cite{Pourovskii05} to solve the effective quantum impurity problem. In this formalism, the self-energy includes  Hartree and Fock diagrams with the bare interaction replaced by the $T$ matrix and  particle-hole contributions with the bare interaction replaced by the particle-hole potential fluctuation matrix. The $T$ matrix and the particle-hole potential fluctuation matrix are in turn expressed in terms of particle-particle and particle-hole susceptibilities. We use the charge self-consistent LDA+DMFT(SPTF) approach  as implemented in an electronic structure code based on a full-potential linear muffin-tin orbital method (LMTO)~\cite{AGrechnev:2007,JMWills:1987,OGranas2012,JMWills2010}.  The LMTO basis sets contain a triple basis for $s$ and $p$ states and a double basis for the $d$ orbitals of YCo$_5$. The basis of the valence electrons is constructed with 4$s$, 4$p$, and 3$d$ states for the Co atoms, and 5$s$, 5$p$, 4$d$ for Y atoms.  The $N_k$ $\mathbf{k}$-points are distributed with the conventional Monkhorst-Pack grid, and the Brillouin zone integration is carried out with the Fermi smearing at a temperature of $T=474\; \text{K}$.  To explore the role of electronic correlation effects arising from the screened Coulomb interaction $U$, we fix the higher order slater integrals of $F^2=7.75\;\text{eV}$, and $F^4=4.85\;\text{eV} $ from Ref.~\onlinecite{OGranas2012}. These values of $F^2$ and $F^4$ for the $d$-orbitals result in a Stoner parameter $J=0.9\;\text{eV}$, which is consistent with the value used in earlier studies of Co metals~\cite{AGrechnev:2007}.  We treat the effect of the Coulomb exchange interaction with $F^2$ and $F^4$ explicitly.

\begin{figure}
\includegraphics[width=1\linewidth]{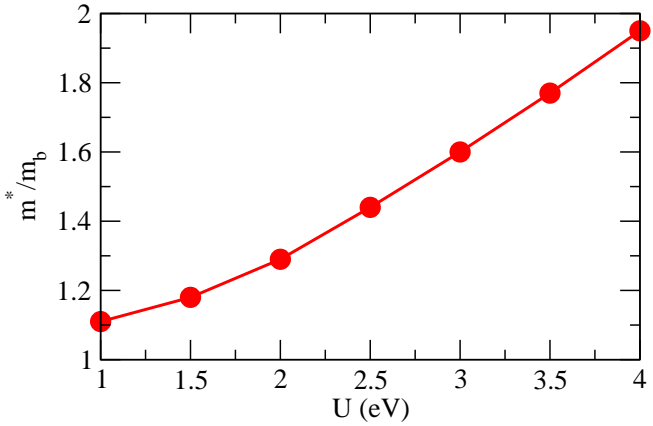}
\caption{\label{Fig2} Ratio between the effective electronic mass, $m^*$, in presence  of on-site Coulomb interaction $U$, and the mass $m_b$ obtained from a LDA calculation. }
\end{figure}

We first explore the relevance of the notion of electronic correlation in the ferromagnetic magnetic metals by studying the quasiparticle renormalization effect. 
In connection with the specific heat coefficient as measured from the thermodynamic experiments, the effective mass enhancement  is proportional to the ratio of the quasiparticle density of states to band one at the Fermi energy: $m^*/m_b=\tilde{\rho}(E_F)/\rho_b(E_F)$. For the cases where the $d$ electrons are active carriers, the band density of states has a predominant  $d$-character: $\rho_b(E_F)= \sum_{i,\alpha} w_i \rho_{b,\alpha}(E_F)$, where $\rho_{b,i,\alpha}$ is the partial density of states at the Fermi energy from the 10 spin orbitals for the $i$-th type of Co atom. Here $\alpha$ is the spin-orbital index, while $w_i$ is the number of equivalent atoms of a given type.  Within a renormalized band theory, we can generalize  the quasiparticle density of states at the Fermi energy as $\tilde{\rho}_b(E_F)= \sum_{i,\alpha} w_i \tilde{\rho}_{b,\alpha}(E_F)$. Here the  spin-orbital dependent quasiparticle density of states at the Fermi energy is given by $\tilde{\rho}_{b,i,\alpha}(E_F)=\rho_{b,\alpha}(E_F)/z_{i,\alpha}$, where the quasiparticle weight is  $z_{i,\alpha}=[1-\partial\text{Im}\Sigma_{\alpha,i}(i\omega_n)/\partial \omega_n|_{\omega_n \rightarrow 0}]^{-1}$ with the self-energy $\Sigma_{\alpha,i}$ defined on the Matsubara frequency  $\omega_n$ axis. 

\begin{figure}[t!]
\includegraphics[width=1\linewidth]{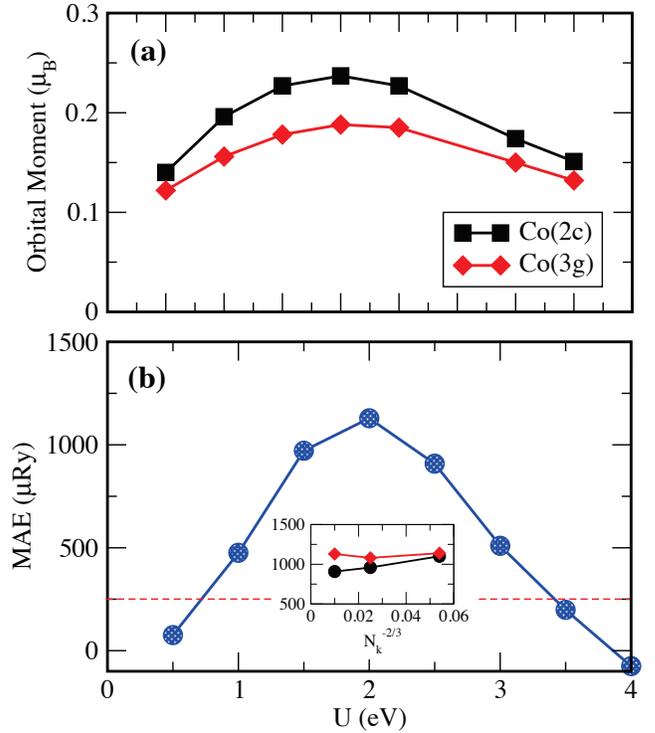}
\label{Fig3} 
\caption{On-site Coulomb $U$ dependence of the orbital magnetic moments on Co sites and  magnetocrystalline anisotropy energy (MAE) per formula unit of YCo$_5$. All the solid curves correspond to the results obtained with the LDA+DMFT method described in the text. The dashed line corresponds to the measured valued according to Ref.~\cite{Skomski99}. The inset to panel (b) shows the MAE dependence  on the number of $\mathbf{k}$ points in the Brillouin zone for two representative values of Hubbard interaction $U=2.0$ eV (red line with diamond symbols) and 2.5 eV (black line with circle symbols). The $\mathbf{k}$-point convergence is reasonably reached.}
\end{figure}

Figure~\ref{Fig2} shows the $U$ dependence of the mass enhancement relative to LDA calculations, $m^*/m_b$, obtained by applying the LDA+DMFT method to YCo$_5$. 
As expected, the effective mass increases monotonically with $U$. The mass enhancement takes values between 1.5 and 2 for $U$ varying between $2.5$ and $4$ eV.
The Sommerfeld coefficient $\gamma$ of the specific heat is proportional to the effective mass of the quasiparticles.  Based on our specific heat measurements of YCo$_5$,  we obtain a Sommerfeld coefficient  $\gamma = 90\; \text{mJ}/\text{mol}\cdot \text{K}^2\cdot \text{f.u.}$, which is $\sim$ 2.7  times larger than the value  $\gamma_b \simeq 33  \; \text{mJ}/\text{mol}\cdot \text{K}^2\cdot \text{f.u.}$ extracted from pure LDA calculations  (see Appendix~\ref{appendix:specificheat}).  
Note that some additional contribution to the electronic renormalization arising from the electron-phonon coupling is not included in our calculations.
Because  $\gamma/\gamma_b$ is equal to  $m^*/m_b$,  this ratio indicates that YCo$_5$ is  a rather correlated metal for $U$ between 2.5 and 4 eV. Values of $U$ in this range have  been previously reported in the YCo$_5$ literature~\cite{AGrechnev:2007,OGranas2012}.

Fig.~\ref{Fig3}(a) shows the orbital magnetic moments on the two inequivalent Co atoms as a function of $U$. The orbital moment of the Co(2c) atoms is always larger than the moment of the Co(3g) atoms and both depend non-monotonically on $U$, reaching their maximum values at $U \simeq 2$ eV.  The results for $U \to 0$  reproduce the values obtained in previous LDA calculations~\cite{Nordstrom92,Takahashi87}, while the moments increase by a factor of $\sim 2$ for $U \simeq 1-3$ eV. This increase  is consistent with our  XMCD measurements, which indicate that the average orbital magnetic moment on the Co ion is 0.20 $\mu_B$.  This observation confirms the relevant role of $U$ on the formation of  a strong orbital moment.

However, the most dramatic effect of the on-site  Coulomb repulsion $U$ appears when we compute the MAE, as is clear from our LDA+DMFT results shown in  Fig.~\ref{Fig3}(b).  By comparing Figs.~\ref{Fig3}(a) and (b), we can see that the MAE and  the size of the orbital moments exhibit the same non-monotonic dependence on $U$. The extrapolated LDA value of the MAE is much lower than the measured value of $K_1 V= 250$ $\mu$Ry shown with a dashed line in Fig.~\ref{Fig3}(b) ($V=0.84 \times 10^{-22}\; \text{cm}^3$ is the volume of the primitive unit cell and $K_1=2.98 \times 10^{18}\;\text{Ry}/\text{cm}^3$~\cite{Skomski99}). {\it However, the MAE increases drastically with  $U$ reaching values that are more than an order of magnitude higher  in the range} $U\sim$1-3.5 eV. This dramatic increase not only explains the reason why LDA calculations systematically  underestimate the MAE of strong magnets, but also  shows the crucial role played by electronic correlations in the development of large magnetic coercivity.  In addition, the MAE obtained from our LDA+DMFT calculations for $U$ between 3 and 3.5 eV is in good agreement with the experimental value.


\section{Conclusions}
The fact that the strongest  magnets are rare-earth based compounds, suggests that the large magnitude of the SO coupling plays a crucial role in the development of high coercivity. One would then expect that the intrinsic magnetocrystalline anisotropy of Nd$_2$Fe$_{14}$B or SmCo$_5$ originates in the crystal field splitting of the rare-earth 4$f$ levels. There is experimental evidence, however, indicating that substantial magnetocrystalline anisotropy
may be associated with the transitional metal sublattice itself. For instance, the coercive field of magnetically hardened Gd$_2$Fe$_{14}$B
is ~2.5 kOe~\cite{Herbst84,Croat84}, but the Gd$^{3+}$  ion has no significant contribution from 4$f$-electrons to the orbital moment, suggesting an increasing role of Gd $5d$-orbital electrons~\cite{MColarieti-Tosti:2003}. In addition, the MAE of SmCo$_5$ is only three times higher than the MAE of YCo$_5$ and Y is non-magnetic. Our results indicate that the MAE of a magnet is dramatically modified by the presence of strong on-site Coulomb interaction $U$ that tends to localize the electrons. We note that enhanced correlations could also be playing a role  in rare-earth based compounds (rare-earths have large ionic radii). This may explain why rare-earth based compounds, in which the rare-earth has no orbital moment, still have  very high MAE. 

By a close comparison of  our LDA+DMFT calculations with  different key experimental measurements, we have shown that  electronic correlation effects play an  essential role in determining the MAE of YCo$_5$. These calculations suggest that the figure of merit of strong magnets can be greatly optimized by tuning the electron Coulomb repulsion $U$.  Our analysis has  natural implications for the search of rare-earth free strong magnets.  While it may be important to retain a large SO coupling, it is equally or even more important to find strongly correlated ferromagnets in order to induce a large enough orbital moment on the transition metal. 
Developing  predictive tools for the MAE of strong magnets is an essential precondition for guiding the search for new materials. Our results indicate that LDA+DMFT techniques  are very promising because they incorporate the relevant  interplay between kinetic and Coulomb energies. Further improvements in impurity solvers should allow to obtain even  more reliable values of the MAE for magnets that are in the intermediate or strong coupling regime (that is, $U$ comparable to or larger than the bandwidth).

\acknowledgments
We are grateful to Tomasz Durakiewicz,  O. Gr\r{a}n\"{a}s, J. Schweizer, F. Tasset, P. Thunstr\"{o}m, and J. M. Wills for helpful discussions. Work at the LANL was performed under the auspices of the U.S.\ DOE contract No.~DE-AC52-06NA25396 through the LDRD program. Part of the theoretical calculations were carried out on a Linux cluster in the Center for Integrated Nanotechnologies, a DOE Office of Basic Energy Sciences user facility.

\appendix
\section{Specific Heat Measurements on YCo$_5$}
\label{appendix:specificheat}
\begin{figure}[h]
\centering
\includegraphics[width=1.0\linewidth,clip]{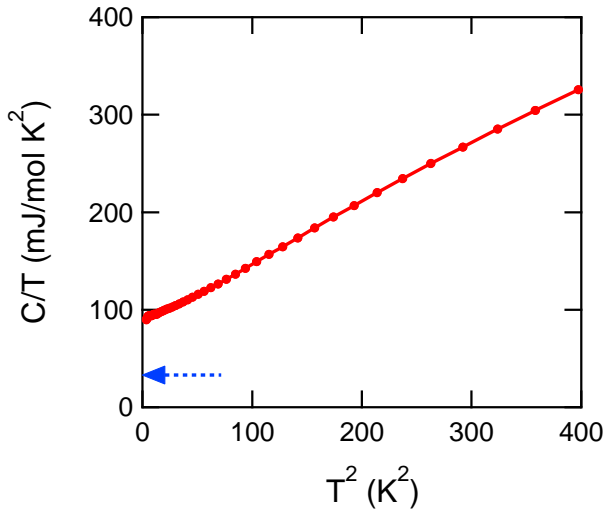}
\caption{(Color online) 
$C/T$ vs $T^2$ for YCo$_5$, from which we obtain the Sommerfeld coefficient $\gamma$  in the 
$T \rightarrow 0$ limit. The dashed arrow is value expected based on the band calculations.
}
\label{Sfig:specific_heat}
\end{figure}

We perform the specific heat measurements on polycrystalline samples of YCo$_5$, which  were made by arc-melting the constituents on a water-cooled copper hearth.
It  was measured down to 2 K in zero magnetic field using a thermal relaxation method implemented in a Quantum Design PPMS-9 device. The data is shown in Fig.~\ref{Sfig:specific_heat}. The Sommerfeld coefficient ($\gamma$) was found to be 90 $\text{mJ}/\text{mol}\cdot \text{K}^2$ by fitting $C/T$ below 10 K to the form of $\gamma + \beta T^2 + \delta T^4$. We attribute $\gamma$ to the electronic  contribution to the heat capacity, while the lattice and magnetic contributions are accounted for by the $\beta T^2$ and $\delta T^4$ terms.

To obtain the mass enhancement due to strong correlations we compare the measured Sommerfeld coefficient to the bare density of states obtained by our DFT calculations using the generalized gradient approximation with the Perdew-Burke-Ernzerhof exchange correlation potential~\cite{PPerdew:96}. Both the full-potential linear muffin-tin orbital method as implemented in the RSPt~\cite{JMWills2010} program and the full-potential linearized augmented plane wave  as implemented in Wien2k~\cite{wien2k} program give the consistent results. 
By summing both spin contributions we find the density of states at the Fermi level $N(E_F) = 14$ states/eV. From this we obtain a mass enhancement $m^*/m_b = \gamma /(\pi^2 k_{B}^{2}N(E_F)/3) = 2.7$.

\section{X-ray Circular Magnetic Dichroism  measurements on YCo$_5$}
The XMCD measurements were carried out in a total electron yield detection scheme at the beam line 4-ID-C of the Advanced Photon Source, Argonne National Laboratory. The beamline 4-ID-C has the ability to generate circularly polarized x-rays at the resonances of 3$d$ elements with high degree of circular polarization ($>$  97\%) by means of an electromagnetic circularly polarizing undulator, including the ability to switch polarization state with a 1 Hz frequency. For the XMCD measurements the samples have been ground into fine powder and been pressed directly into electrically conducting carbon tape and placed in contact with a Cu holder. The Cu holder was electrically isolated from the cold finger by a sapphire disk. The samples were placed into a 7 Tesla superconducting magnet with a variable temperature insert. All scans were carried out at a temperature $T$ = 20 K and over an energy range of 770 to 810 eV to measure the Co $L_3$- and $L_2$-edges (778.1 and 793.2 eV, respectively). Total electron yield data sets $\mu^+$ and $\mu^-$ recorded with left- and right-circularly polarized x-rays, respectively, were background subtracted and edge-step normalized (edge is normalized to one). Moreover, each measurement was carried out for magnetic fields $H$ $=$ 6~Tesla directed along and opposite to the photon wave vector, respectively, to check for experimental artifacts. Using $\mu^+$ and $\mu^-$ the normalized XANES ($\mu_0=\frac{\mu^++\mu^-}{2}$) and XMCD ($\Delta\mu=\mu^+-\mu^-$) data sets for YCo$_5$ were obtained. The orbital contribution to the magnetic moment was then extracted using the sum rules for $3d$ transition metals \cite{Thole92, Carra93}:

\begin{equation}\label{eq:angular}
L= -\frac{4}{3}n_h \frac{\Delta I_{L_3}+\Delta I_{L_2}}{I_{L_3}+I_{L_2}},
\end{equation}
where $n_h$ is the number of holes in the $3d$ shell, and $n_h$~$=$ 3  for the 3$d^7$ configuration of Co in YCo$_5$. $I_{L_2}$/$I_{L_3}$ are the integrated intensity in the isotropic white lines at the $L_2$/$L_3$ edges, and $\Delta I_{L_2}$/$\Delta I_{L_3}$ are the integrated intensities in the partial dichroic signal.


\end{document}